          [2001/04/25 1.1 (PWD)] 
\documentclass[a4paper,twocolumn]{esapub2005} 
\def\kte{kT_{\rm e}} 
\def\taut{\tau_{\rm T}}

\def\be{\begin{equation}} 
\def\ee{\end{equation}} 
\def\beq{\begin{eqnarray}} 
\def\eeq{\end{eqnarray}} 
\def\gax {\ifmmode{_>\atop^{\sim}}\else{${_>\atop^{\sim}}$}\fi}  
\def\IGR{IGR~J00291+5934}
\def\SAX{SAX~J1808.4-3658}
\def\J1751{XTE~J1751-305}
\def\hete{HETE~J1900.1-2455}
\def\1807{XTE~J1807-294}

\usepackage{times} 
\usepackage{natbib} 
\usepackage{graphicx}
 
\title{Accreting X-ray millisecond pulsars observed with INTEGRAL} 
\author{Maurizio Falanga} 
\affil{CEA Saclay, DSM/DAPNIA/Service d'Astrophysique, F-91191, Gif
sur Yvette, France; e-mail: mfalanga@cea.fr} 
\affil{Unit\'e mixte de recherche Astroparticule et  
Cosmologie, 11 place Berthelot, 75005 Paris, France}

\begin{document} 
 
 
\maketitle 
 
\begin{abstract} 
I review the properties of three X-ray accreting millisecond pulsars
observed with INTEGRAL.  Out of
seven recently discovered accretion-powered pulsars (one discovered by
INTEGRAL), three were observed with the INTEGRAL satellite up to 300 keV. 
Detailed timing and spectral results will be
presented, including data obtained during the most recent outburst of
the pulsar HETE J1900.1-2455. Accreting X-ray millisecond pulsars are key
systems to understand the 
spin and accretion history of neutron stars. They are also a good
laboratory in which to study the source spectra, pulse profile, and
phase shift between 
X-ray pulses in different energy ranges which give additional information
of the X-ray production processes and emission environment.
\end{abstract} 
 
\section{Introduction} 

Many low-mass X-ray binaries (LMXB) consist of a neutron star
accreting from a low mass companion star ($<$ 1 $M_{\odot}$).  
Accreting matter may spin up the neutron
star (NS); therefore, one of the possible endpoints of
the evolution of a low-mass X-ray binary is expected to be a
millisecond pulsar, i.e. a
rapidly spinning NS with a rather weak, $\sim 10^{8}$ Gauss, surface
magnetic field. Although evidence for  rapidly spinning NS in LMXBs
was inferred from the burst oscillations that were seen during type I
X-ray bursts in several systems \citep[see][for a review]{stroh01},
the detection of millisecond pulsations in persistent emission remained
elusive for many years until the discovery of the first accreting
millisecond pulsar by \citet{wvdk98}. Since that
time, a total of seven accreting MSPs have been detected.

All of the accreting MSPs are X-ray transients; they spend most of
the time in a quiescent phase, with X-ray luminosities of order
of $10^{31}-10^{33}$ erg s$^{-1}$. They sometimes show X-ray outbursts
reaching X-ray luminosities of $10^{36}-10^{37}$ erg s$^{-1}$,
during which coherent pulsations are observed with
frequencies in the range between 180 and 435 Hz  \citep[see reviews
by][]{w05,p06}. This frequency is interpreted as the NS rotation
frequency given by a hot spot (or spots) in an atmospheric layer of
the rotating NS \citep{Chakrabarty03}. MSPs represent a new
class of objects connecting  accretion powered X-ray pulsars with
 rotation powered millisecond radio pulsars.  

MSP  energy spectra are successfully fitted   by a two-component model 
consisting of a multicolor blackbody soft X-ray emission and a 
Comptonized spectrum for the hard X-ray emission. 
The soft thermal component could be associated with the 
radiation from the accretion disc and/or the heated NS surface around the 
shock \citep[see e.g.][]{gp05,p06}. The hard emission is likely to be
produced by thermal Comptonization in the hot accretion shock on the
NS surface \citep{gdb02,pg03} with the seed photons coming from the
stellar surface. The observed hard spectra are similar to the spectra
observed from atoll sources in their hard, low-luminosity state \citep{b00}.
A representative spectrum of the three observed MSP (XTE~ J1807-294
\citep{mfa05}, IGR~J00291+5934  \citep{mfb05}, and HETE J1900.1-2455
\citep{mf06}) with INTEGRAL is shown in Figure 1, 2 and 3.  

Recent years have shown INTEGRAL gave additional contributions to the
study of MSPs. Our timing analysis of
IGR~J00291+5934 showed for the first time that INTEGRAL/ISGRI is
capable of detecting  the pulse profile of a 1.67 ms pulsar up to 150
keV. This allowed us to study the pulsed fraction up to this high energy. We
confirmed for the first time the increase of the pulsed fraction with
energy in an accretion powered MSP, explained by the action of the
Doppler effect on the exponentially cutoff Comptonization spectrum
from the hot spot or from a componization model where we account the
pulsed fraction to be produced in a corona cloud. We also measured
soft time lags with a complex energy dependence. Similarly to SAX
J1808.4-3658 and XTE~J1751-350, the time lags increase rapidly with
energy until ~10 keV. However, in IGR~J00291+5934, the
time lags first increase and then decrease slightly, saturating above
15 keV, and possibly reaching zero around 50 keV. This is a intriguing
result, and if confirmed or observed in another object could be a
serious challenge to any time-lag model. Additionally, with RXTE we
measured an increase of the NS spin period during accretion for the
first time. This provided a strong
confirmation of the theory of 'recycled' pulsars in which the old
neutron stars in LMXBs become millisecond radio pulsars through
spin-up by transfer of angular momentum by the accreting material
\citep{mfb05}. 

Analysis of the recent ToO observation of HETE~J1900.1-2455 is also
discussed. We observe in this object Type I X-ray bursts showing,
at high energy, evidence of photospheric radius expansion in the burst
profile. Using JEM-X/ISGRI data at high energy, and assuming the bolometric
burst peak luminosity during photospheric radius expansion to be
saturated at the Eddington limit, we measure
the source distance to be around 5 kpc. A very interesting result is
that the pulsations at the  spin frequency are not always observed in
this source. 

\section{X-ray spectra}

\begin{figure}[ht]
\centering 
\includegraphics[width=0.65\linewidth,angle=-90]{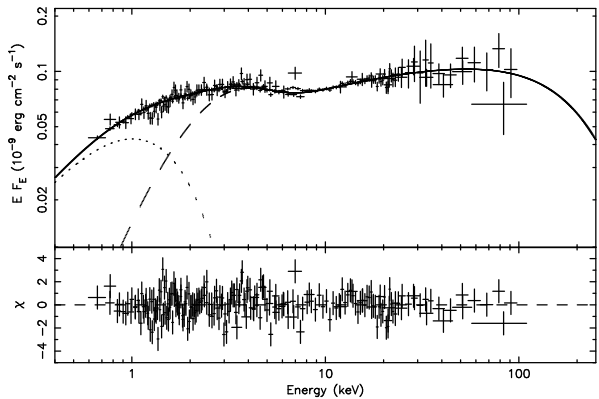} 
\caption{Simultaneous INTEGRAL, XMM-Newton and RXTE 
        spectra of XTE~J1807--294 fitted with an absorbed disk black body, {\sc
	diskbb}, plus {\sc compps} model. The EPIC-pn and MOS2 spectra in the 
        0.5--10 keV range and PCA/HEXTE in the 3--200 keV and
	IBIS/ISGRI in the 20--200 keV range are shown. The {\sc
	diskbb} model is shown by a dotted curve, the dashed curve
	gives the {\sc compps} model, while the total spectrum is
	shown by a solid curve. The lower panel presents the residuals.} 
\label{fig:XTEJ1807} 
\end{figure} 
 
\begin{figure}[ht] 
\centering 
\includegraphics[width=0.65\linewidth,angle=-90]{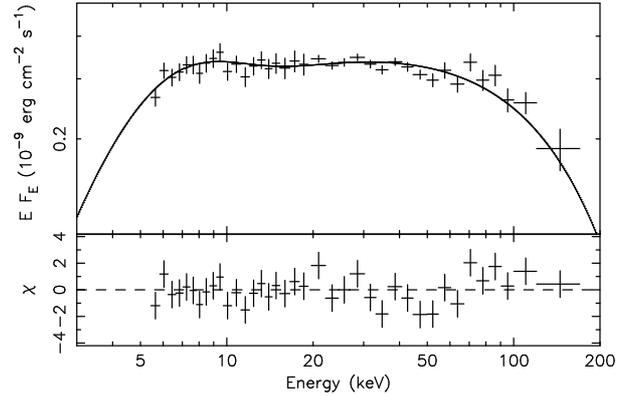} 
\caption{The unfolded spectrum of \IGR\ fitted with an absorbed
	{\sc compps} model. The data points correspond to the JEM-X
	(5--20 keV)  and ISGRI (20--200 keV) spectra,
	respectively. The total spectrum  of the model is shown by a
	solid curve. The lower panel presents the residuals between
	the data and the model.}  
\label{fig:IGRJ00201}
\end{figure} 
 
\begin{figure}[ht] 
\centering 
\includegraphics[width=0.65\linewidth,angle=-90]{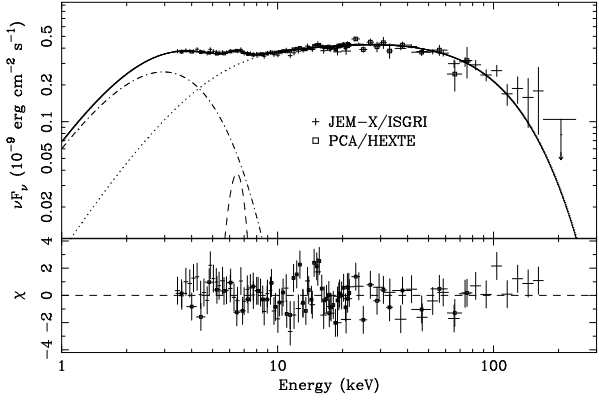} 
\caption{The unfolded spectrum of HETE~J1900.1--2455 fitted with a
	{\sc compps} model plus a {\sc bb} and a gaussian line. The
	data points correspond to the PCA (3--22 keV), JEM-X (3--22
	keV), HEXTE (16--90 keV) and the ISGRI (20--300 keV) spectra,
	respectively. The blackbody model is shown by a  dot-dashed
	curve, the dotted curve gives the {\sc compps} model, the
	dashed curve is the line, the total spectrum is shown by the
	solid curve. The lower panel presents the residuals between
	the data and the model. 
}
\label{fig:HeteJ1900} 
\end{figure} 
 
For the INTEGRAL observed MSP sources the high energy spectrum could be
described with a photon index of $\sim$ 1.8--2.0 and a cut-off energy at $\sim$
80--100 keV. However, the model does not describe the spectrum below
15 keV well, which requires a more complex description. 
MSP  energy broad-band spectra are successfully fitted by a
two-component model consisting of a multicolor blackbody soft X-ray
emission and a  
Comptonized spectrum, for the hard X-ray emission. The soft thermal
component could be associated with the radiation from the accretion
disc and/or the heated NS surface around the shock. The hard emission
is likely to be produced by thermal Comptonization in the hot
accretion shock on the NS surface with the seed photons coming from the
stellar surface.  The spectra are best fitted with the thermal
Comptonization model {\sc compps} in the slab geometry 
\citep{ps96}. The main model parameters are the Thomson optical depth 
$\tau_{\rm T}$ across the slab, the electron temperature 
$kT_{\rm  e}$, the soft seed photon temperature $kT_{\rm seed}$, and  
the inclination angle $\theta$ between the slab normal and the line of sight. 
The seed photons are supposed to be injected from the bottom of the
slab. The soft thermal emission is fitted by a simple blackbody, {\sc bb}, 
or a multi-temperature disc blackbody, {\sc dbb}, model \citep{mitsuda84}.  
The best fit parameters for the three observed MSP with INTEGRAL are
reported in Table \ref{table:spec}. In Fig. \ref{fig:XTEJ1807},
\ref{fig:IGRJ00201},  and \ref{fig:HeteJ1900}, we show the unfolded spectrum 
and the residuals of the data to the {\sc bb} or {\sc dbb} plus 
{\sc compps} model.  

The spectra of \1807, \IGR, and \hete\ are well described by 
a combination of thermal Comptonization  and a disk black body or
simple  black body. 
The hard spectral component contributes most of the observed flux (70--80
per cent), even though a soft component (disk black body) is required
by the data. If the accretion disk is truncated at the 
magnetosphere radius where the material can accrete along the magnetic 
field towards the pole of a NS, the matter over the 
pole is heated by a shock to a temperature of $\sim$ 30--50 keV. The hot spot 
at the surface with temperature $\sim$ 0.8--1.5 keV gives rise to the seed 
photons for Comptonization in the hot plasma. Since the hard X-ray 
emission is pulsed (see Sec. 3.1), a fraction of it must 
originate from the regions confined by the magnetic field. The most 
obvious source of hard X-rays is the place where material collimated by 
the magnetic fields impacts  the surface. 
 
For \1807, if we consider the inclination of the system to be
$60^{\circ}<i<83^{\circ}$, this allows us to determine the inner disk radius
which lies  in the range 20--40 km (for the distance of 8 kpc). 
The apparent area of the seed
photons is $A_{\rm seed} \sim 26 (D/8 \ {\rm kpc})^{2} {\rm km}^{2}$,
which is also what expected from a hot spot on a NS surface. 
Many characteristics of the source are similar to those observed in
other MSPs, SAX~J1808.4--3658 \citep{gdb02,pg03} and XTE~J1751--305
\citep{gp05}. For \IGR\ no additional thermal-like
component (either blackbody or disk blackbody) was required by the
fit, very likely because the bulk of its emission occurs below 3 keV,
outside the covered energy range. However, the apparent area of the seed 
  photons, which turns out to be 
$A_{\rm seed}\sim 21 (D/5\ {\rm kpc})^{2}$ km$^2$, could correspond to
a  hot spot radius of $\sim2.5$ km during the INTEGRAL ToO outburst phase.  

Only \hete's spectrum differs from other MSP in requiring thermal 
soft X-ray emission with nearly double the temperature. 
We could not distinguish between the multi-temperature blackbody and 
single blackbody models, as both gave  comparable parameters and $\chi^2$.   
However, for a distance of 5 kpc,  the disk blackbody gives 
an  inner disk radius, $R_{\rm in} \sqrt{\cos\,i} = 2.6$ km, 
smaller than the expected NS radius. For the blackbody emission, the
fit implies the apparent area of the emission region  
$A_{\rm seed} \sim 14$ km$^2$, which could be consistent  with the  
heated NS surface around the accretion shock \citep{pg03,gp05}.
From our spectral fits we infer that this emission is not
likely produced in a multi-temperature accretion 
disk but  more likely arises from thermal emission at the NS
surface.

We note that the product $\taut\times\kte$ is very close 
in the three sources. The spectral shape and the product $\taut\times\kte$ 
are very stable during the outburst as is observed 
in other sources too \citep[e.g.][]{gp05}. The constancy of the
spectral slopes  during the outbursts and their extreme similarity 
in different MSPs can be used as an argument that the 
emission region geometry does not depend on the accretion rate. 
If the  energy dissipation takes place in a hot shock, while the cooling 
of the electrons (that emit X/$\gamma$-rays via thermal Comptonization) 
is determined by the reprocessing of the hard 
X-ray radiation at the neutron star surface (so called two-phase model, 
the spectral slope is determined by the energy balance in the hot phase 
and  is a function of the geometry. At constant geometry (e.g. slab), 
the temperature depends on the optical depth, but $\taut \times \kte$ 
is approximately constant. 
 
\begin{table*}[htb] 
\begin{center} 
\caption{\label{table:spec}Best-fit spectral parameters with  
 {\sc compps} + {\sc bb} (or {\sc diskbb}) model.}   
\begin{tabular}{llll} 
\hline 
\noalign{\smallskip}  
 & XTE~J1708-294 &  IGR~J00291+5934& HETE~J1900.1-2455 \\ 
\hline 
\noalign{\smallskip}  
$N_{\rm H} (10^{22} {\rm cm}^{-2})$ & $0.56$ (f) & $0.28$ (f) & $0.16$ (f)\\ 
$kT_{\rm in}$ or $kT_{\rm BB}$(keV) & -- & 0.43$^{+0.04}_{-0.04}$ & 
0.8$^{+0.02}_{-0.02}$\\ 
$R_{\rm in} \sqrt{\cos i}^{\ a}$ (km) & $13.4^{+2.2}_{-1.3}$ & -- & --\\ 
$R_{\rm bb}^{\ a}$ (km)      & -- & -- & $4.8^{+0.7}_{-0.6}$\\ 
$kT_{\rm e}$ (keV) & $37.2^{+28}_{-10}$ & $49^{+2}_{-6}$ & $27.9^{+1.8}_{-1.4}$\\ 
$kT_{\rm seed}$ (keV)    & $0.75^{+0.04}_{-0.04}$ &
$1.49^{+0.16}_{-0.32}$ & $1.4^{+0.16}_{-0.32}$\\  
$\tau_{\rm T}$ & $1.7^{+0.5}_{-0.8}$ & $1.12^{+0.04}_{-0.07}$ & $2.0^{+0.06}_{-0.1}$\\ 
$A_{\rm seed}^{a}$ (km$^2$)  & $26^{+23}_{-12}$  & 
$20.7^{+12.6}_{-4.5}$ & $14.2^{0.3}_{0.3}$\\ 
cos $\theta $  & $0.79^{+0.07}_{-0.06}$  & $0.6^{+0.06}_{-0.09}$  &
0.59$^{+0.05}_{-0.07}$ \\ 
$L_{\rm bol}^{a}$ ($10^{36}$ erg s$^{-1}$) & 3.6 & 3.7 & 4.9\\ 
\noalign{\smallskip}  
\hline  
\noalign{\smallskip}  
\end{tabular}  
\end{center} 
\end{table*}

\section{Timing}

The timing analysis of
IGR~J00291+5934 showed for the first time that INTEGRAL/ISGRI is
capable of detecting  the pulse profile of a 1.67 ms pulsar up to 150
keV. This allowed us to study the pulsed fraction up to this high energy. We
confirmed for the first time the increase of the pulsed fraction with
energy in an accretion powered MSP. This can be explained by two
models:  the action of the
Doppler effect on the exponentially cutoff Comptonization spectrum
from the hot spot, or  a componization model where  the
pulsed fraction is  produced in a corona cloud. We also measured
soft time lags with a complex energy dependence. Similarly to SAX
J1808.4-3658 and XTE~ J1751-350, the time lags increase rapidly with
energy until ~10 keV. However, in IGR~~J00291+5934, the
time lags first increase and then decrease slightly, saturating above
15 keV, and possibly reaching zero around 50 keV, see Fig. 6.

\subsection{Pulsed fraction of X-rays}

\begin{figure}[ht] 
\centering 
\includegraphics[width=0.53\linewidth,angle=90]{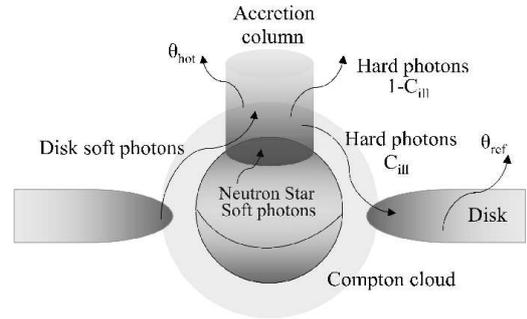} 
\caption{This cartoon illustrates the different emission patterns 
  responsible for the time lags of the pulsed emission.  
  $\theta_{\rm e}^{\rm hot}$ and $\theta_{\rm e}^{\rm ref}$  are the
  dimensionless temperature of the accretion column and reflector,
  respectively, and $C_{\rm ill}\sim 0.1$ is the disk illumination fraction. 
Soft time lag of the pulsed emission is the result of downscattering
  of hard X-ray photons in the relative cold plasma of the disk. 
A fraction of hard X-ray photons $1-C_{\rm ill}$  directly seen by the
  observer are  upscattered soft  photons coming from NS and  the
  disk. We account for the pulsed flux (pulsed fraction) as being
  produced in the  corona cloud.}
\label{fig:geom} 
\end{figure} 

We found for \IGR\ using the RXTE and INTEGRAL/ISGRI data that the
pulsed fraction gradually increases with energy  
from $\sim 6\%$ at 6 keV to $\sim 12-20 \%$ near 100 keV. This is the 
first time such behaviour has been measured for any of the
accretion-powered MSPs.  
 
Since MSPs are rapidly rotating, a first interpretation is given 
through a Doppler boosting model which affects the fast rotating spot
emitting  pattern as a black body.  
When the spot moves towards the observer, the emission 
increases, while for a spot moving away, the flux drops. 
The Doppler factor reaches the maximum a quarter of a period before the 
peak of the projected area,  shifting the emission peak towards an
earlier phase. The observed flux due to the Doppler effect varies 
as the Doppler factor to the power $(3+\Gamma)$ 
 \citep{pg03,vp04}, where the photon index $\Gamma$ could be a function 
of energy. If the Doppler factor varies around 1 with 2\% amplitude, 
we get 10\% variability. 
 
\begin{figure}[ht] 
\centering 
\includegraphics[width=0.9\linewidth]{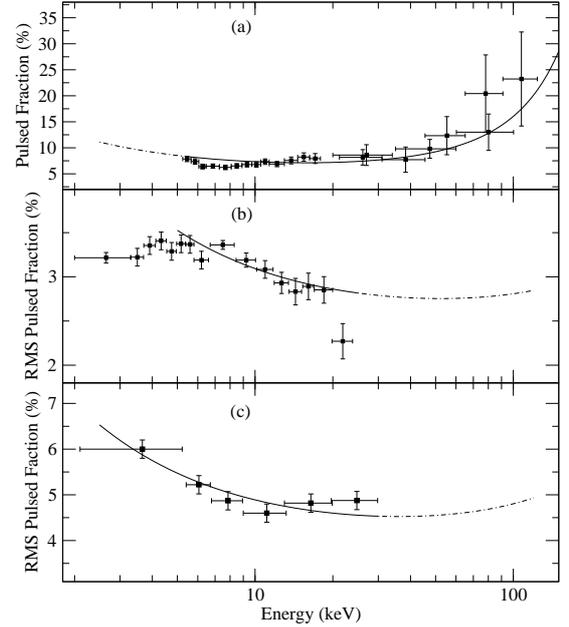} 
\caption{The observed  energy dependent pulsed fraction of
(a) \IGR\, (b) XTE J1751-305, and (c) SAX J1808.4-3658  along with the
  best-fit comtonization model. 
}
\label{fig:PF} 
\end{figure}

The Comptonized spectrum can be approximated as 
\be 
F_E \propto E^{-(\Gamma_0-1)} \exp\left( -[E/E_{\rm c}]^{\beta}\right) , 
\ee 
where  $E_{\rm c}\sim \kte$ is the energy of the cutoff  and parameter 
$\beta\sim2$ describes its sharpness. 
The local photon index is then 
\be 
\Gamma(E)\equiv 1 - \frac{\mbox{d} \ln F_E }{\mbox{d} \ln E} = 
\Gamma_0 + \beta (E/E_{\rm c})^{\beta} . 
\ee 
At low energies, $\Gamma\approx\Gamma_0$, and 
rms (or pulsed fraction) is a very weak function of energy. 
Close to the cutoff, the spectral index rapidly increases and 
the pulsed fraction should grows with energy, as observed (see 
Fig.~\ref{fig:PF}a). 
The Comptonization models predict  softening of the total spectrum 
with simultaneous hardening of the pulsed spectrum at higher energies.

On the other hand, the pulsed fraction of \IGR\ vs
energy can be also explained by the energy 
dependent  electron cross-section $\sigma_e(E) =\sigma_{\rm  T}(1-2z)$ 
and consequently by Compton cloud  optical depth as a function energy   
\begin{equation}
\tau_{\rm cl}(E)=\tau_{{\rm T, cl}}(1-2z),
\label{electron_tau}
\end{equation}
where $z=E/m_ec^2$ is a dimensionless photon energy.
In Figure \ref{fig:PF} we show the energy dependent pulsed fraction of
(a) \IGR, (b) \J1751\ and (c) \SAX.
In fact, we assume that the accretion column is embedded in
Compton cloud  as shown in Figure \ref{fig:geom}  \citep[see also][for
 the geometry details]{TCW}.   
Because $\tau_{\rm cl}(E)$ decreases with energy,  at higher
energies a larger fraction of the pulsed direct hard X-ray radiation   
\begin{equation}
A_{\rm rms,es}(E)=A_{\rm rms,es}(0)\exp[-\tau_{\rm cl}(E)]
\label{A-fraction}
\end{equation}
originating  in the accretion column can escape to the observer. 

A different scenario of the energy dependent amplitude  formation
would be if 
one can assume that there is no electron (Compton) cloud between the
accretion column (where the Comptonization spectrum is formed) and the
observer, but the energy dependence of the amplitude is formed as a
result of   upscattering.    
The soft photons upscattered  off of hot electrons (of the accretion
column)   increase their energy  with a number of scatterings. 
On the other hand, the amplitude  of the pulsed radiation exponentially
decreases 
with a number of scatterings $A_{\rm rms}(u)=A_{\rm rms}(0)\exp(-\beta u)$,
where $\beta$ is the inverse of an average number of scatterings
\citep[see e.g.][]{ST80} and  consequently with a energy of the
upscattered photon 
\begin{equation}
A_{\rm rms,up}[u(E)]=A_{\rm rms,up}(E_s)(E/E_s)^{-\alpha_{\rm cl}}
\label{A-upscatter_fraction}
\end{equation}
 (where $\alpha_{\rm cl}=\beta/4\theta$ is approximately the energy  index
 of the Comptonization spectrum).  A characteristic seed (disk and NS)
 photon energy  is $E_s<  5$ keV  and consequently,  all photons at
 higher energies ($>5$ keV)  are produced by upscattering.  The
 amplitude  of upscattered photons  (in which the energy is higher 5 keV)
 should decrease with energy.     

In the general case,  some fraction of the pulsed  soft photons,
$A_{\rm rms,up}$,  upscatter  off hot electrons in the accretion column on the
way out as another fraction of the pulsed photons, $A_{\rm rms,es}$, forming
the hard X-ray tail escape to the observer passing through the Compton cloud.  
In this  case a following combination of Eqs. (\ref{A-fraction}
and \ref{A-upscatter_fraction}) 
\begin{equation}
A_{\rm rms}(E)= A_{\rm rms,up}(E/E_s)^{-\alpha_{\rm cl}} + A_{\rm rms,es}e^{-\tau_{\rm cl}(E)} 
\label{A-general_fraction}
\end{equation}
leads us to the formula of the emergent pulsed amplitude.

In Figure \ref{fig:PF}  we present  the best-fit model (see
Eq. \ref{A-general_fraction})  along with data points.  We  fit only
the data points which correspond to  energies higher than the seed
photon energy.   We found for (a) \IGR\ the Compton cloud 
optical depth to be $\tau_{\rm T, cl} = 3\pm0.2$ and $A_{\rm
  rms,es} = 98^{+2}_{-6}$ \%,  $A_{\rm rms,up} = 5.2\pm0.9$ \%. For (b)
  $\tau_{\rm T, cl} = 0.25\pm0.09$ and $A_{\rm 
  rms,es} = 3.5\pm0.8$ \%,  $A_{\rm rms,up} = 1.8\pm1.1$ \% and (c)
    $\tau_{\rm T, cl} = 0.4\pm0.2$ and $A_{\rm rms,es} = 5.9\pm6$ \%,
      $A_{\rm rms,up} = 2.5\pm1.2$ \%, respectively. For all of the fits
      $\alpha_{\rm cl}$ was $\sim0.8$.

\subsection{Time lag}

\begin{figure}[ht] 
\centering 
\includegraphics[width=0.9\linewidth]{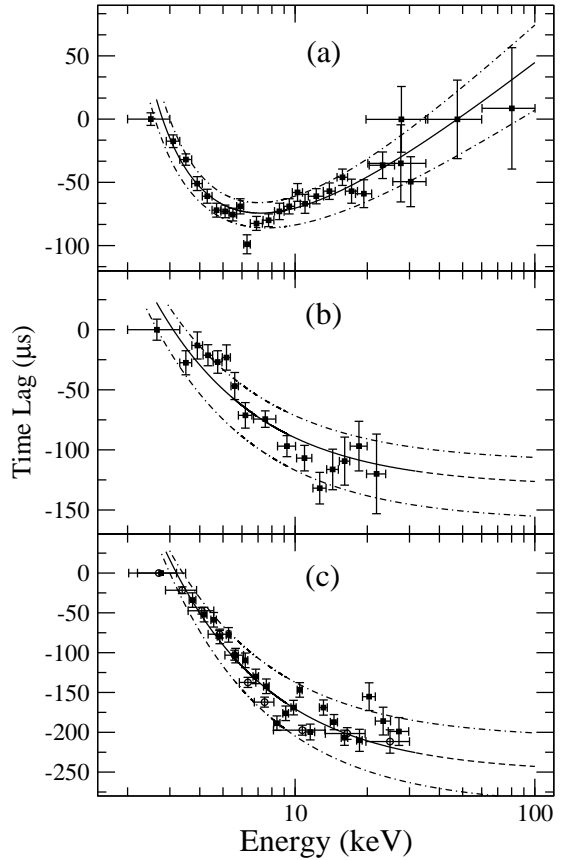} 
\caption{The measured soft time lag of the pulse profile versus energy
 (crosses) with respect to the first energy channel. The
 best-fit curve using the Comptonization model (see Eq. 7)
 is shown with the solid line.   The dot-dashed lines in panel (a)  correspond to the upper and lower limits of the electron number densities of the
 Comptonization emission area, $n_{\rm e}^{\rm hot}$ and  of  the
 reflector, $n_{\rm e}^{\rm ref}$   in \IGR.
   The panels (b) and (c)  are related  to \J1751\ and (c) for \SAX.
 The dot-dashed lines correspond to the upper and lower
 limits of  $n_{\rm e}^{\rm ref}$.
}
\label{fig:TL} 
\end{figure} 

Additional information for the X-ray production processes and
emission environment can be obtained by studying the pulse profile
and phase shift between X-ray pulses at different energy ranges. For
\SAX\ and later for \J1751\ it was found that the low-energy pulses
lag behind the high-energy pulses (soft phase/time lags) monotonically
increasing with energy  and saturating at about 10--20 keV
\citep{CMT98,Ford00,gp05}.

First, this phase/time lag effect was interpreted  as a result
of the photon delay due to downscattering of  hard X-ray photons in the
relatively cold plasma of the disk or NS surface \citep{CMT98,TCW}.   
They argued that the photon time lags were an intrinsic signature of
interaction of the Comptonized radiation with the NS and accretion disk
plasma. Moreover, the absolute values of time lags (about hundreds
$\mu$s) are consistent with the electron scattering time scale $t_{\rm
  C}=\tau_{\rm T}(L/c)$.  The effective Thomson  optical depth of the cold
reflector $\tau_{\rm T}=n_e\sigma_{\rm T}L$ is about a few and
typical sizes of the NS photosphere and half-width of the disk $L$ are
of the 
order of $10^{6}$ cm. On the other hand, \citet{pg03} suggested that
the lags may be produced by a combination of different angular
distribution of the radiation components  and relativistic effects.
The observed spectrum of \IGR\ consists of a black body 
from the neutron star surface and a component produced by 
Comptonization of these seed photons in the hot electron  region, 
presumably a shock which can be represented as a plane-parallel slab. 
The angular distributions of the black body and Comptonized 
photons emitted by the slab are significantly different. 
The difference in the emission patterns causes the two components 
to show a different variability pattern as a function of the pulsar phase, 
with the hard Comptonized component leading the soft black body 
component. However, the  soft lags found  up to $\sim100$ keV in the 1.67 ms
accreting MSP \IGR, show a more complex energy dependence \citep{mfa05}. 

Recently, \citet{MFLT06} derived a  Comptonization model for the
observed properties of the energy dependent soft/hard time lags. This
accounts for  the soft lag by downscattering of hard
X-ray photons in the relatively cold plasma of the disk or neutron
star surface. A fraction of soft X-ray photons 
coming from  the disk or neutron star surface are upscattered off hot
electrons of accretion column  and this effect leads  to the hard lags
as a result of thermal Comptonization of the  soft photons. The model
allows  the observed soft and hard 
lags  to be reproduced by the downscattered and upscattered radiation
as a function of 
the electron number densities of the reflector, $n_{e}^{\rm ref}$, and
the accretion column,  $n_{ e}^{\rm hot}$. In the case of the
accretion-powered millisecond pulsars \IGR, \J1751,  and \SAX\ the
observed time lags agree very strongly with the model (see
Fig. \ref{fig:TL}).  The resulting time lags are seen as a linear
combination of the 
positive (hard) time lags formed in the Comptonization emission area and
the negative (soft) ones formed in accretion disk and NS photosphere as a
result of reflection of the hard radiation (see Fig. \ref{fig:geom}): 
$$
\Delta t  =  -\frac{C_{\rm ill}}{\sigma_{\rm T}n_{\rm e}^{\rm ref} c} \times
$$
\begin{equation}
\biggl[\frac{1}{4\theta_{\rm ref}}\ln
\frac{1-4\theta_{\rm
      ref}/z}{1-4\theta_{\rm ref}/z_{*}}
      -\frac{n_{\rm
  e}^{\rm ref}}{n_{\rm e}^{\rm hot}}\frac{1-C_{\rm ill}}{C_{\rm ill}}
\frac{\ln(z/z_{*})}{\ln[1+(3+\alpha)\theta_{\rm hot}]}\Biggr]
\label{res_timelag}
\end{equation}
where $n_{\rm e}^{\rm ref}$ is the electron number density of the reflector,
$n_{ e}^{\rm hot}$ is the electron number density of the
Comptonization emission area 
(accretion column) and $\theta_{\rm ref}=kT_e^{\rm ref}/(m_{e} c^2)$ is a
dimensionless temperature of the reflector.  
We assume a typical value of $\theta_{\rm ref} < 0.7$ keV/511 keV.   
 $kT_{\rm e}^{\rm hot}$ and $\alpha$  are the best-fit parameters for the hot
plasma temperature and spectral index of the  Comptonization spectrum,
and $\theta_{\rm  hot}=kT_{ e}^{\rm hot}/(m_{ e} c^2)$. 

We also assume that the seed photon energy is near 
the lowest energy of the downscattered photons, i.e. $z_s\approx z_{\ast}$.   
An observational value of $E_{\ast}$  is  about  3 keV, and we
assume that the  value of  the illumination factor $C_{\rm ill}$ is
about 0.1 \citep[see for details][]{MFLT06}. To examine the predicted
time lag using the Comptonization model 
Eq. (\ref{res_timelag}), we use the observed pulse phase lag data
and best-fit spectral parameters  $kT_{\rm e}^{\rm hot}$  and
$\alpha$ of the accreting MSPs \IGR, \J1751, and \SAX. The \SAX\ data
were taken from \citet{CMT98,Ford00} and represented in
Figure \ref{fig:TL} by open circles and filled squares, respectively. The
data for \J1751 and  \IGR\ were taken by  \citet{gp05,mfa05}. For the reflector
temperature we used  $kT_{\rm e}^{\rm ref} = 0.4$ keV for all  fits.
The  model to fit the time lags in \J1751\ and \SAX\ 
has only one free parameter (see \citep{MFLT06}), the number
density of the ``cold'' reflector $n_{\rm e}^{\rm ref}$. Presumably, in \J1751\
and \SAX\ the density of the Comptonization region is much higher than
that of the ``cold'' reflector, i.e. $n_{\rm e}^{\rm hot}\gg n_{\rm
  e}^{\rm ref}$. It is not surprising that the seed photons are
Comptonized (upscattered) in the very dense plasma of the accretion column
\citep[see e.g.][]{BS76,BW06}. The best-fit values are $n_{\rm e}^{\rm
  ref}= 6.3\times10^{19}$ cm$^{-3}$ and $3.2\times10^{19}$ cm$^{-3}$
  for \J1751\ and \SAX, respectively. The fits of the time lag
data for \IGR\ provide us the best-fit values of the ``cold'' and hot
plasma  densities. Both the positive (upscattering) and negative
(downscattering) time lags  contribute to the apparent time lags
because $n_{\rm e}^{\rm ref}= 6.9\times10^{18}$ cm$^{-3}$ and $n_{\rm
  e}^{\rm hot}=  2.1\times10^{18}$ cm$^{-3}$ are of the same order of
magnitude.     

The time lag  data of  \IGR, \J1751, and \SAX\ are collected in
different time intervals lasting from hours to days
\citep{CMT98,Ford00,gp05,mfa05}. However, a hydrodynamical 
(density perturbation) time,  $t_{hydro}$, in the
innermost part of X-ray NS source is on the order of the ratio of the
NS radius to the sound speed, namely  $t_{hydro}\sim R_{\rm NS}/c_{\rm
sound}\gax 0.1$ s. 
Thus, during the data collection periods the densities of the
surrounding plasma can vary. As a result, the time lags also
change because they are very sensitive to density variations (see
Eq. \ref{res_timelag}). In fact,  these density variations can be inferred
from the time lag data. One can make a band between two curves of time
lags vs energy which contain all of the time lag data points (see
Fig. \ref{fig:TL}). This allows  constraints to be put on $n_{\rm
  e}^{\rm ref}$ and $n_{\rm e}^{\rm hot}$.  
The lower and upper curves in Figure \ref{fig:TL}
correspond (a) to $n_{\rm e}^{\rm ref}=  (6.1-8.0)\times10^{18}$ cm$^{-3}$
and  $n_{\rm e}^{\rm hot} = (1.6-2.6)\times10^{18}$ cm$^{-3}$, respectively.
For the sources (b) \J1751\ and (c) \SAX\, the density variations are 
$n_{\rm e}^{\rm ref}= (6.0-6.6)\times10^{19}$ cm$^{-3}$ and  $n_{\rm
e}^{\rm ref}= (2.9-3.6)\times10^{19}$ cm$^{-3}$.   
Thus, the plasma density of the ``cold'' reflector can change as much as
10 \% during the entire data collection.

Soft lags are observed only if $n_e^{ref}\ll n_e^{hot}$ and thus the
relative fraction of the dowscattering time lags in the total time lag
sum is $\sim$100\%.   
 However, Comptonization in a non-uniform accretion medium might
 account for the observed time lag as a non-monotonic function of energy. 
Using this model,  upper and lower limits of the
atmosphere density (density variation) in the region of phase/time lag
measurements were determined. Using the observed energy dependent
pulsed amplitude one can find the variation of the Thomson optical depth of
Compton cloud in which the accretion column is embedded.

\section{Bursts}

\begin{figure}[ht] 
\centering 
\includegraphics[width=1.3\linewidth,angle=-90]{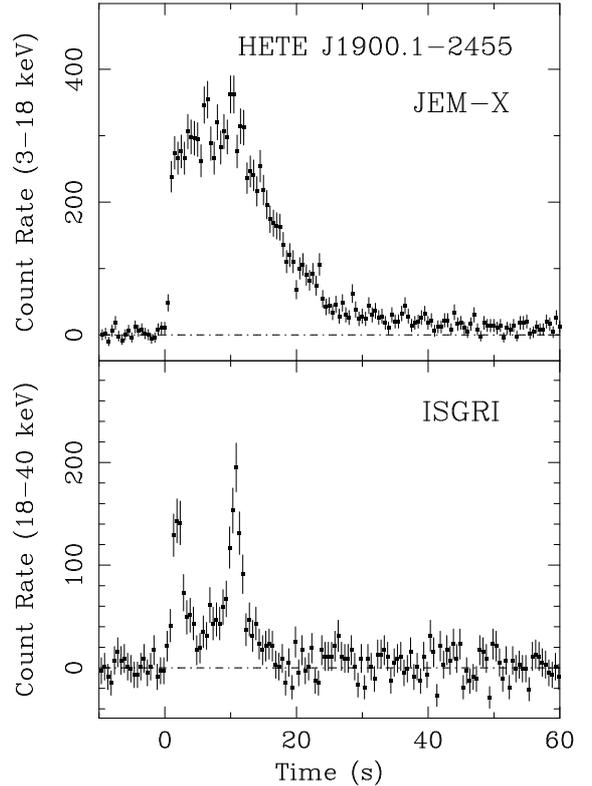} 
\caption{A bright X-ray burst detected from HETE~J1900.1--2455. The JEM-X 
  (3--20 keV; upper panel) and IBIS/ISGRI (18--40 keV; lower panel) 
  net light curves are shown (background subtracted). The time bin is 
  0.5 s for both IBIS/ISGRI and JEM-X light curve. At high energy the 
  burst shows strong evidence of photospheric radius expansion.}
\label{fig:lc_burst}
\end{figure}

MSP are also known to exhibit Type-I X-ray bursts. These bursts were
observed for \SAX\ and XTE~J1814-338 \citep{gmh06,gc06}. Several X-ray bursts
have also been observed  for \hete\ by various observatories
\citep{gmm06}. One of these bursts was also observed  during the
INTEGRAL ToO observation \citep{mf06}. In Figure \ref{fig:lc_burst} we 
show the JEM-X and ISGRI burst light curve (28 October 2005, 10:25:12
UTC) in different energy bands. The burst rise time was $0.23\pm0.05$ 
s. The double peak profile is clearly evident at 
high energy (lower panel) within the first 12 s, while  
during this time the intensity at lower energy (upper panel) remains
constant.  This can be interpreted as a consequence of a photospheric
radius  expansion (PRE) episode during the  first part of the outburst
\citep[see  e.g.,][]{vh95}. When a burst undergoes a PRE episode, 
the luminosity remains nearly constant at 
the Eddington value, the atmosphere expands, and its temperature  
decreases resulting in  a double-peak profile  observed at high
energies. The tail of the burst at high energy can be seen for about 5
s after the PRE episode.  
 
A time-resolved analysis of the net burst spectrum based on the 
JEM-X/ISGRI  3-50 keV energy band data was well fit by a 
photoelectrically absorbed blackbody. During the first 12 s, the unabsorbed 
bolometric flux was almost constant at $F_{\rm peak}=9.5(2)\times10^{-8}$ erg 
cm$^{-2}$ s$^{-1}$, while the blackbody temperature dropped in the middle, 
simultaneously with an increase by a factor of $\sim1.5$ in blackbody 
radius. The observed temperature reached a peak at $\sim$ 2.5 keV, and then 
gradually decreased. The softening of the emission towards the end of the 
decay phase is also indicated by the e-folding decay times 
of $12.5\pm0.5$ s in the 3--6 keV to $4.3\pm0.7$ s in the 12--20 keV
energy band.  This behavior is typically observed during PRE X-ray bursts.
 
The burst fluence was $f_{\rm b}=1.67(6)\times10^{-8}$erg cm$^{-2}$,
calculated by integrating the measured $F_{\rm bol,bb}$ over the burst
duration of $\sim 50$ s. The effective burst duration was $\tau=f_{\rm
b}/F_{\rm peak}=18.2(8)$ s, and the ratio of the observed persistent flux to
the net peak flux was $\gamma=F_{\rm pers}/F_{\rm  peak}=0.021(1)$.  
The burst has the same spectral parameters as previous 
bursts for this source observed with HETE-2 or RXTE \citep{gmh06}. 
Assuming a helium burst at the Eddington limit  and canonical 
NS parameters (1.4 solar mass and radius of 10 km),  the source
distance is estimated to be $\sim5$ kpc.  

From the observed {\it INTEGRAL} burst properties and the  mass
accretion rates  inferred  from  
the persistent luminosity, the present theory predicts that this
burst is pure helium burning. For helium 
flashes, the fuel burns rapidly, since there are no slow weak 
interactions, and the local Eddington limit is often exceeded. These 
conditions lead to PRE bursts with a duration, set mostly by the time 
it takes the heat to escape, of the order of 5--10 s, as 
observed. In the framework of the 
thermonuclear-flash models \citep[e.g.,][]{vh95} the burst 
duration, $\tau < 20$ s, and the ratio of 
the observed persistent flux to the net peak flux $\gamma\approx0.02$ indicate
a hydrogen-poor burst which is in agreement with the companion star being a 
helium-rich brown dwarf \citep{kaaret06}.

Because there were no other bursts observed during the {\em INTEGRAL}
observation, the burst recurrence time, $\Delta t_{rec}$,  must be at
least one day. We can compute the ratio of the total energy 
emitted in the persistent flux to that emitted in  the burst 
$$
\alpha= \frac{F_{pers}}{f_b} \Delta t_{rec} = \frac{\gamma}{\tau}
\Delta t_{rec}  =   \frac{0.021}{18.2} \Delta t_{rec}  > 100, 
$$
which is consistent with  pure helium bursts \citep[see e.g.][]{vh95}. 
Taking  the burst total energy release $E_{b}=5\times 10^{39}$ erg   
(derived from the fluence $f_b$)  
and  He burning efficiency of $\epsilon_{\rm He}\approx 1.7$ MeV/nucleon $\approx 1.6\times10^{18}$
erg g$^{-1}$, we estimated the amount of fuel burned during the burst 
$E_{b}/\epsilon_{\rm He}\sim3.1\times10^{21}$ g. 
For the mean mass accretion rate of 2 per cent of the Eddington,
a burst recurrence time of 2.2 days is expected.

\section{Conclusions} 
\label{sec:conclusions}

The science of accreting millisecond pulsars has been carried forward
by the high-energy capabilities of INTEGRAL in concert with the 
timing resolution capabilities of RXTE and the  spectral
resolution of XMM at low energies. INTEGRAL observations have 
contributed to our ability to characterize the high energy spectrum of these
sources, and to study their interesting time-lag and pulse fraction
behavior over a wide energy range, up to $\sim$150 keV. 

The reason for the lack of coherent pulsations in the persistent emission
from LMXBs has been a longstanding open question in X-ray
astronomy. For the first time, we are able to observe a system (\hete) known
to contain a millisecond period pulsar, which has continued in
outburst after its pulsations have disappeared. 
The  transition of \hete\ from an X-ray
millisecond pulsar to a persistent LMXB could indicate that
there is a population of suppressed  X-ray millisecond pulsars among the
non-pulsating LMXBs. Detailed observations of this source at the epoch
of pulsation supression can help to solving the long-standing issue of
missing pulsations in persistent LMXB emission.

\section*{Acknowledgments} 

I am grateful to the INTEGRAL MSP paper collaborators J. Poutanen,
E. W. Bonning, L. Kuiper, J. M. Bonnet-Bidaud, A. Goldwurm,
W. Hermsen, L. Stella, and L. Titarchuk on the time/phase lag paper.

\end{document}